\documentclass[a4paper]{jpconf}
\usepackage{graphicx}

\usepackage{textcomp}
\usepackage{amsmath}
\usepackage{verbatim}

\def\kpn{K^+\rightarrow\pi^+\nu\bar\nu}
\def\klpn{K_{L}\rightarrow\pi^0\nu\bar\nu}

\begin{document}

\title{Kaon physics overview}

\author{Jason Aebischer}

\address{Physik-Institut, Universit\"at Z\"urich, CH-8057 Z\"urich, Switzerland}

\ead{jason.aebischer@physik.uzh.ch}

\begin{abstract}
The present status of Kaon physics is summarized from a theory point of view. Focus is laid on $\Delta S=2$, $\Delta S=1$ as well as on rare Kaon decays, where progress has been made for instance in the case of $\varepsilon_K$ and $K_S\to \mu^+\mu^-$. Furthermore, several New Physics scenarios are discussed in the context of rare Kaon decays and other Kaon observables.
\end{abstract}

\section{Introduction}
Kaon Physics is an excellent probe for various CKM elements, the study of CP violation as well as New Physics (NP) effects of the Standard Model (SM). It has therefore gained a lot of attention recently, from the experimental \cite{Goudzovski:2022scl} as well as from the theory side \cite{Aebischer:2022vky}. In the following we will present the latest theory results in Kaon physics, where we lay special focus on $\Delta S=2,1$ as well as on rare decays. Finally we discuss several NP scenarios that can accommodate the current data.

\section{Kaon mixing}

Starting with $\Delta S=2$ processes,
the mass difference of the two neutral Kaons, $\Delta M_K$,
is measured to be \cite{Workman:2022ynf}
\begin{equation}
  (\Delta M_K)_{\text{exp}}=3.484(6)\times 10^{-15} \text{GeV}\,.
\end{equation}

From the theory side $\Delta M_K$ is known up to NNLO \cite{Brod:2011ty}. Long distance (LD) contributions have been estimated in the Dual QCD (DQCD) approach \cite{Bijnens:1990mz,Buras:2014maa} and from Lattice QCD \cite{Bai:2014cva,Christ:2015pwa,Bai:2018mdv}, leading to a SM prediction of \cite{Wang:2018csg}
\begin{equation}
  (\Delta M_K)_{\text{SM}}=7.7(1.3)(2.1)\times 10^{-15} \text{GeV}\,.
\end{equation}

The SM prediction is therefore slightly above the experimental value and hence motivates NP scenarios that suppress the SM value. We will discuss possible scenarios in Sec.~\ref{sec:NP}.

The other important observable in Kaon mixing is the mixing parameter $\varepsilon_K$, which is measured to be \cite{Workman:2022ynf}
\begin{equation}
  |\varepsilon_K|_{\text{exp}}=2.228(11)\times  10^{-3}\,.
\end{equation}

Again, NNLO corrections are known \cite{Brod:2011ty,Brod:2010mj,Brod:2021qvc,Herrlich:1996vf,Brod:2022har} and also LD contributions have been estimated \cite{Buras:2008nn,Buras:2010pza, RBC:2020kdj}.
Large progress has been made recently in the work of Brod et al. \cite{Brod:2019rzc}, in which the unitarity relation of the CKM matrix has been used in a special manner to reduce the uncertainty from the charm-quark contribution, leading to
\begin{equation}
  |\varepsilon_K|_{\text{SM}}=2.16(6)(8)(15)\times  10^{-3}\,.
\end{equation}

This novel approach allowed to reduce the uncertainty by one order of magnitude and therefore renders $\varepsilon_K$ a precision observable, up to parametric uncertainties dominantly represented by a strong $V_{cb}$ dependence.

\section{$\varepsilon^\prime/\varepsilon$ and $\Delta I=1/2$}

Other interesting processes in the Kaon sector are $\Delta S=1$ transitions.
The most prominent one being the ratio measuring direct over indirect CP violation, denoted by  $\varepsilon^\prime/\varepsilon$.
Also here NLO \cite{Buras:1991jm,Buras:1992tc,Buras:1992zv,Buras:1993dy,Ciuchini:1992tj,Ciuchini:1993vr} as well as NNLO results are available \cite{Buras:1999st,Cerda-Sevilla:2016yzo,Cerda-Sevilla:2018hjk}. The LD effects have been estimated in different approaches such as Lattice QCD \cite{Blum:2015ywa,RBC:2015gro}, Chiral perturbation theory ($\chi$PT) \cite{Gisbert:2017vvj} as well as DQCD \cite{Buras:2015xba}, leading to different theory predictions
\begin{align}
  (\varepsilon^\prime/\varepsilon)_{\text{Lattice}} &= (21.7 \pm 8.4)\times 10^{-4} \,, \quad \cite{RBC:2020kdj}\\
  (\varepsilon^\prime/\varepsilon)_{\chi\text{PT}}&=(14\pm 5)\times 10^{-4}\,, \quad \cite{Cirigliano:2019cpi}\\
  (\varepsilon^\prime/\varepsilon)_{\text{DQCD}}&=(5\pm 2)\times 10^{-4}\,, \quad \cite{Buras:2020wyv}\\
  (\varepsilon^\prime/\varepsilon)_{\text{Lattice+IB}}&= (13.9\pm 5.2)\times 10^{-4}\,, \quad \cite{Aebischer:2020jto}
\end{align}
where for the last one the computation was perfomred using Lattice results together with isospin-breaking (IB) corrections \cite{Cirigliano:2003gt,Buras:2020pjp} and NNLO QCD effects \cite{Aebischer:2019mtr}. More effort from the theory side is needed in order to reduce the uncertainties and therefore provide a clear cut theory prediction.

$\varepsilon^\prime/\varepsilon$ has been measured most precisely by NA48 \cite{NA48:2002tmj} and KTeV \cite{KTeV:2002qqy}, leading to an averaged value of
\begin{equation}\label{eq:epe_exp}
  (\varepsilon^\prime/\varepsilon)_{\text{exp}}=(16.6 \pm 2.3) \cdot 10^{-4}\,.
\end{equation}

All of the theory predictions except for the DQCD result are compatible with the experimental value in Eq.~\eqref{eq:epe_exp}, although there is still room left for NP effects, which will be discussed briefly in Sec.~\ref{sec:NP}.

Another important $\Delta S=1$ transition is the $\Delta I =1/2$ rule, defined by the ratio $R$ of the Isospin 0 and 2 amplitudes:
\begin{equation}
  R\equiv{\rm Re}(A_0)/{\rm Re}(A_2)\,.
\end{equation}

Experimentally, this ratio is rather large, signaling dominance of the $I=0$ over the $I=2$ amplitude. One finds \cite{Workman:2022ynf}
\begin{equation}
  {\rm Re}(A_0)_{\text{exp}}=27.04(1) \cdot 10^{-8}\,\text{GeV}\,,\quad {\rm Re}(A_2)_{\text{exp}}=1.210(2) \cdot 10^{-8}\,\text{GeV}
\end{equation}

leading to a ratio of
\begin{equation}
  R_{\rm exp} = 22.35(4)\,.
\end{equation}

From the theory side the ratio has been estimated using Lattice and DQCD techniques, leading to
\begin{align}
  R_{\text{Lattice}} &= 19.9(2.3)(4.4) \,, \quad \cite{RBC:2020kdj}\\
  R_{\text{DQCD}}&= 16(2)\,. \quad \cite{Buras:2014maa}
\end{align}

Also here more effort from the theory side is needed in order to clarify the situation.

\section{Rare Kaon Decays}
One of the most important players in Kaon Physics are rare Kaon decays, such as $K\to \pi \nu\bar\nu$ and $K\to (\pi)\ell^+\ell^-$ transitions. These decays are sensitive to various combinations of CKM elements and are theoretically very clean.

Starting with the $K\to \pi \nu\nu$ decays, from the experimental side we have for $\kpn$ from NA62 \cite{NA62:2021zjw}
\begin{equation}
  \mathcal{B}(\kpn)_\text{exp}=(10.6^{+4.0}_{-3.5}\pm 0.9)\times 10^{-11}\,.
\end{equation}

For $K_L\to \pi^0\nu\nu$ only an upper bound from KOTO \cite{KOTO:2018dsc} is available:
\begin{equation}
  \mathcal{B}(\klpn)_\text{exp}\le 3.0\times 10^{-9}\,.
\end{equation}

Both decays are known up to NLO \cite{Buchalla:1993bv,Buchalla:1993wq,Misiak:1999yg} and NNLO \cite{Buras:2005gr,Buras:2006gb,Gorbahn:2004my}. Also Isospin breaking \cite{Mescia:2007kn} as well as non-perturbative effects \cite{Isidori:2005xm} have been estimated. One down-side of the two decays is their strong dependence on them CKM element $V_{cb}$, which lowers their potential to probe NP effects. The problem is related to the disagreement between the inclusive \cite{Bordone:2021oof} and exclusive \cite{Aoki:2021kgd} determination of $V_{cb}$. In order to overcome this issue it has been proposed recently by Buras and Venturini to consider suitable ratios of observables, in which the dependence on $V_{cb}$ drops out \cite{Buras:2021nns,Buras:2022nfn,Buras:2022qip}. This method allowed to produce the most precise predictions of the $K\to \pi\nu\bar\nu$ branching ratios \cite{Buras:2022wpw}:
\begin{align}
  \mathcal{B}(K^+\to \pi^+\nu\bar\nu)&=(8.60\pm 0.42)\times 10^{-11} \,,\\
  \mathcal{B}(K_L\to \pi^0\nu\bar\nu)&=(2.94\pm 0.15)\times 10^{-11}\,.
\end{align}

Other interesting rare Kaon decays are $K_S\to\mu^+\mu^-$, $K_L\to\pi^0e^+e^-$ and $K_L\to\pi^0\mu^+\mu^-$.

From the experimental side only upper bounds from LHCb \cite{LHCb:2020ycd} and KTeV \cite{KTeV:2003sls,KTEV:2000ngj} are available:
\begin{align}
  \mathcal{B}(K_S\to\mu^+\mu^-)_{\rm LHCb} &< 2.1 \times 10^{-10} \,,\\
  \mathcal{B}(K_L\to\pi^0e^+e^-)_{\rm KTeV} &< 28 \times 10^{-11} \,,\\
  \mathcal{B}(K_L\to\pi^0\mu^+\mu^-)_{\rm KTeV} &< 38 \times 10^{-11} \,.
\end{align}
The theory predictions for these processes are given by:
\begin{align}
  \mathcal{B}(K_S\to\mu^+\mu^-)_{\rm SM} &= (5.2\pm 1.5)\times 10^{-12} \,,\quad \cite{Isidori:2003ts,DAmbrosio:2017klp}\\
  \mathcal{B}(K_L\to\pi^0e^+e^-)_{\rm SM} &= 3.54^{+0.98}_{-0.85}(1.56^{+0.62}_{-0.49})\times 10^{-11} \,,\quad \cite{Mescia:2006jd}\\
  \mathcal{B}(K_L\to\pi^0\mu^+\mu^-)_{\rm SM} &= 1.41^{+0.28}_{-0.26}(0.95^{+0.22}_{-0.21})\times 10^{-11} \,,\quad \cite{Mescia:2006jd}
\end{align}

Recently, progress has been made concerning the process $K_S\to \mu^+\mu^-$ by Dery et. al~\cite{Dery:2021mct}, where the authors showed, that $\mathcal{B}(K_S\to \mu^+\mu^-)$ can be determined experimentally by relating it to $K_L\to \mu^+\mu^-$. This allows to predict the quantity $|V_{ts}V_{td}\sin(\beta+\beta_s)|$ up to the percent level, rendering $K_S\to \mu^+\mu^-$ another precision observable.

\section{New Physics}\label{sec:NP}
Several of the above mentioned observables have been studied in the context of effective field theories, such as the SM Effective Field Theory (SMEFT) or the Weak Effective Theory (WET). A
SMEFT analysis of $\Delta M_K$ and other $\Delta F=2$ observables can be found for instance in \cite{Aebischer:2020dsw}.

Master formulas for $\varepsilon^\prime/\varepsilon$ in the SMEFT and WET have been derived at LO in \cite{Aebischer:2018csl,Aebischer:2018quc,Aebischer:2018rrz} and even at NLO in \cite{Aebischer:2021hws}.

Correlations between the rare Kaon decays and $\varepsilon^\prime/\varepsilon$ as well as lepton flavour violating decays have been studied in $Z^\prime$ \cite{Aebischer:2019blw}.
Finally, Rare Kaon decays and $\varepsilon^\prime/\varepsilon$ together with $\Delta M_K$ and $\varepsilon_K$ as well as their correlations have been studied in the context of the SMEFT and flavour-violating $Z^\prime$ couplings in \cite{Aebischer:2020mkv}. In this context the effect of back-rotation \cite{Aebischer:2020lsx} in the QCD penguin scenario for $\varepsilon^\prime/\varepsilon$ has been ruled out.

Furthermore, in the context of UV completions the constraints from Kaon mixing as well as $\mathcal{B}(\kpn)$ and $\mathcal{B}(\klpn)$ have been studied in the 4321 model in \cite{Crosas:2022quq}.

Another interesting way to study NP effects is to examine $q^2$-distributions for the $K\to\pi\nu\nu$ decays. The NP contributions to these decays have been computed for vector operators in \cite{Li:2019fhz} and scalar degrees of freedom in \cite{Deppisch:2020oyx}. Such $q^2$-distributions, although experimentally hard to access might be available in the coming decade.

\section{Acknowledgments}
I thank the organizers for their kind invitation. Furthermore I thank Andrzej Buras, Gino Isidori and Zachary Polonsky for useful discussions. This project has received funding from the European Research Council (ERC) under the European Union's Horizon 2020 research and innovation programme under grant agreement 833280 (FLAY), and by the Swiss National Science Foundation (SNF) under contract 200020\_204428.


\section*{References}
\begin{thebibliography}{99}

\bibitem{Goudzovski:2022scl}
E.~Goudzovski, E.~Passemar, J.~Aebischer, S.~Banerjee, D.~Bryman, A.~Buras, V.~Cirigliano, N.~Christ, A.~Dery and F.~Dettori, \textit{et al.}
[arXiv:2209.07156 [hep-ex]].

\bibitem{Aebischer:2022vky}
J.~Aebischer, A.~J.~Buras and J.~Kumar,
[arXiv:2203.09524 [hep-ph]].

\bibitem{Workman:2022ynf}
R.~L.~Workman \textit{et al.} [Particle Data Group],
PTEP \textbf{2022}, 083C01 (2022)
doi:10.1093/ptep/ptac097

\bibitem{Brod:2011ty}
J.~Brod and M.~Gorbahn,
Phys. Rev. Lett. \textbf{108}, 121801 (2012)
doi:10.1103/PhysRevLett.108.121801
[arXiv:1108.2036 [hep-ph]].

\bibitem{Brod:2010mj}
J.~Brod and M.~Gorbahn,
Phys. Rev. D \textbf{82}, 094026 (2010)
doi:10.1103/PhysRevD.82.094026
[arXiv:1007.0684 [hep-ph]].


\bibitem{Bijnens:1990mz}
J.~Bijnens, J.~M.~Gerard and G.~Klein,
Phys. Lett. B \textbf{257}, 191-195 (1991)
doi:10.1016/0370-2693(91)90880-Y

\bibitem{Buras:2014maa}
A.~J.~Buras, J.~M.~G\'erard and W.~A.~Bardeen,
Eur. Phys. J. C \textbf{74}, 2871 (2014)
doi:10.1140/epjc/s10052-014-2871-x
[arXiv:1401.1385 [hep-ph]].

\bibitem{Bai:2014cva}
Z.~Bai, N.~H.~Christ, T.~Izubuchi, C.~T.~Sachrajda, A.~Soni and J.~Yu,
Phys. Rev. Lett. \textbf{113}, 112003 (2014)
doi:10.1103/PhysRevLett.113.112003
[arXiv:1406.0916 [hep-lat]].

\bibitem{Christ:2015pwa}
N.~H.~Christ, X.~Feng, G.~Martinelli and C.~T.~Sachrajda,
Phys. Rev. D \textbf{91}, no.11, 114510 (2015)
doi:10.1103/PhysRevD.91.114510
[arXiv:1504.01170 [hep-lat]].

\bibitem{Bai:2018mdv}
Z.~Bai, N.~H.~Christ and C.~T.~Sachrajda,
EPJ Web Conf. \textbf{175}, 13017 (2018)
doi:10.1051/epjconf/201817513017

\bibitem{Wang:2018csg}
B.~Wang,
PoS \textbf{LATTICE2018}, 286 (2019)
doi:10.22323/1.334.0286
[arXiv:1812.05302 [hep-lat]].

\bibitem{Brod:2021qvc}
J.~Brod, S.~Kvedarait\.{e} and Z.~Polonsky,
JHEP \textbf{12}, 198 (2021)
doi:10.1007/JHEP12(2021)198
[arXiv:2108.00017 [hep-ph]].

\bibitem{Herrlich:1996vf}
S.~Herrlich and U.~Nierste,
Nucl. Phys. B \textbf{476}, 27-88 (1996)
doi:10.1016/0550-3213(96)00324-0
[arXiv:hep-ph/9604330 [hep-ph]].

\bibitem{Brod:2022har}
J.~Brod, S.~Kvedaraite, Z.~Polonsky and A.~Youssef,
[arXiv:2207.07669 [hep-ph]].

\bibitem{Buras:2008nn}
A.~J.~Buras and D.~Guadagnoli,
Phys. Rev. D \textbf{78}, 033005 (2008)
doi:10.1103/PhysRevD.78.033005
[arXiv:0805.3887 [hep-ph]].

\bibitem{Buras:2010pza}
A.~J.~Buras, D.~Guadagnoli and G.~Isidori,
Phys. Lett. B \textbf{688}, 309-313 (2010)
doi:10.1016/j.physletb.2010.04.017
[arXiv:1002.3612 [hep-ph]].

\bibitem{RBC:2020kdj}
R.~Abbott \textit{et al.} [RBC and UKQCD],
Phys. Rev. D \textbf{102}, no.5, 054509 (2020)
doi:10.1103/PhysRevD.102.054509
[arXiv:2004.09440 [hep-lat]].

\bibitem{Brod:2019rzc}
J.~Brod, M.~Gorbahn and E.~Stamou,
Phys. Rev. Lett. \textbf{125}, no.17, 171803 (2020)
doi:10.1103/PhysRevLett.125.171803
[arXiv:1911.06822 [hep-ph]].

\bibitem{Buras:1991jm}
A.~J.~Buras, M.~Jamin, M.~E.~Lautenbacher and P.~H.~Weisz,
Nucl. Phys. B \textbf{370}, 69-104 (1992)
doi:10.1016/0550-3213(92)90345-C

\bibitem{Buras:1992tc}
A.~J.~Buras, M.~Jamin, M.~E.~Lautenbacher and P.~H.~Weisz,
Nucl. Phys. B \textbf{400}, 37-74 (1993)
doi:10.1016/0550-3213(93)90397-8
[arXiv:hep-ph/9211304 [hep-ph]].

\bibitem{Buras:1992zv}
A.~J.~Buras, M.~Jamin and M.~E.~Lautenbacher,
Nucl. Phys. B \textbf{400}, 75-102 (1993)
doi:10.1016/0550-3213(93)90398-9
[arXiv:hep-ph/9211321 [hep-ph]].

\bibitem{Buras:1993dy}
A.~J.~Buras, M.~Jamin and M.~E.~Lautenbacher,
Nucl. Phys. B \textbf{408}, 209-285 (1993)
doi:10.1016/0550-3213(93)90535-W
[arXiv:hep-ph/9303284 [hep-ph]].

\bibitem{Ciuchini:1992tj}
M.~Ciuchini, E.~Franco, G.~Martinelli and L.~Reina,
Phys. Lett. B \textbf{301}, 263-271 (1993)
doi:10.1016/0370-2693(93)90699-I
[arXiv:hep-ph/9212203 [hep-ph]].


\bibitem{Ciuchini:1993vr}
M.~Ciuchini, E.~Franco, G.~Martinelli and L.~Reina,
Nucl. Phys. B \textbf{415}, 403-462 (1994)
doi:10.1016/0550-3213(94)90118-X
[arXiv:hep-ph/9304257 [hep-ph]].

\bibitem{Buras:1999st}
A.~J.~Buras, P.~Gambino and U.~A.~Haisch,
Nucl. Phys. B \textbf{570}, 117-154 (2000)
doi:10.1016/S0550-3213(99)00810-X
[arXiv:hep-ph/9911250 [hep-ph]].

\bibitem{Cerda-Sevilla:2016yzo}
M.~Cerd\`a-Sevilla, M.~Gorbahn, S.~J\"ager and A.~Kokulu,
J. Phys. Conf. Ser. \textbf{800}, no.1, 012008 (2017)
doi:10.1088/1742-6596/800/1/012008
[arXiv:1611.08276 [hep-ph]].

\bibitem{Cerda-Sevilla:2018hjk}
M.~Cerd\`a-Sevilla,
Acta Phys. Polon. B \textbf{49}, 1087-1096 (2018)
doi:10.5506/APhysPolB.49.1087

\bibitem{Blum:2015ywa}
T.~Blum, P.~A.~Boyle, N.~H.~Christ, J.~Frison, N.~Garron, T.~Janowski, C.~Jung, C.~Kelly, C.~Lehner and A.~Lytle, \textit{et al.}
Phys. Rev. D \textbf{91}, no.7, 074502 (2015)
doi:10.1103/PhysRevD.91.074502
[arXiv:1502.00263 [hep-lat]].

\bibitem{RBC:2015gro}
Z.~Bai \textit{et al.} [RBC and UKQCD],
Phys. Rev. Lett. \textbf{115}, no.21, 212001 (2015)
doi:10.1103/PhysRevLett.115.212001
[arXiv:1505.07863 [hep-lat]].

\bibitem{Gisbert:2017vvj}
H.~Gisbert and A.~Pich,
Rept. Prog. Phys. \textbf{81}, no.7, 076201 (2018)
doi:10.1088/1361-6633/aac18e
[arXiv:1712.06147 [hep-ph]].

\bibitem{Buras:2015xba}
A.~J.~Buras and J.~M.~G\'erard,
JHEP \textbf{12}, 008 (2015)
doi:10.1007/JHEP12(2015)008
[arXiv:1507.06326 [hep-ph]].

\bibitem{Cirigliano:2019cpi}
V.~Cirigliano, H.~Gisbert, A.~Pich and A.~Rodr\'\i{}guez-S\'anchez,
JHEP \textbf{02}, 032 (2020)
doi:10.1007/JHEP02(2020)032
[arXiv:1911.01359 [hep-ph]].

\bibitem{Buras:2020wyv}
A.~J.~Buras,
Acta Phys. Polon. B \textbf{52}, no.1, 7-41 (2021)
doi:10.5506/APhysPolB.52.7
[arXiv:2101.00020 [hep-ph]].

\bibitem{Aebischer:2020jto}
J.~Aebischer, C.~Bobeth and A.~J.~Buras,
Eur. Phys. J. C \textbf{80}, no.8, 705 (2020)
doi:10.1140/epjc/s10052-020-8267-1
[arXiv:2005.05978 [hep-ph]].

\bibitem{Cirigliano:2003gt}
V.~Cirigliano, G.~Ecker, H.~Neufeld and A.~Pich,
Eur. Phys. J. C \textbf{33}, 369-396 (2004)
doi:10.1140/epjc/s2003-01579-3
[arXiv:hep-ph/0310351 [hep-ph]].

\bibitem{Buras:2020pjp}
A.~J.~Buras and J.~M.~G\'erard,
Eur. Phys. J. C \textbf{80}, no.8, 701 (2020)
doi:10.1140/epjc/s10052-020-8299-6
[arXiv:2005.08976 [hep-ph]].

\bibitem{Aebischer:2019mtr}
J.~Aebischer, C.~Bobeth and A.~J.~Buras,
Eur. Phys. J. C \textbf{80}, no.1, 1 (2020)
doi:10.1140/epjc/s10052-019-7549-y
[arXiv:1909.05610 [hep-ph]].

\bibitem{NA48:2002tmj}
J.~R.~Batley \textit{et al.} [NA48],
Phys. Lett. B \textbf{544}, 97-112 (2002)
doi:10.1016/S0370-2693(02)02476-0
[arXiv:hep-ex/0208009 [hep-ex]].

\bibitem{KTeV:2002qqy}
A.~Alavi-Harati \textit{et al.} [KTeV],
Phys. Rev. D \textbf{67}, 012005 (2003)
[erratum: Phys. Rev. D \textbf{70}, 079904 (2004)]
doi:10.1103/PhysRevD.70.079904
[arXiv:hep-ex/0208007 [hep-ex]].

\bibitem{NA62:2021zjw}
E.~Cortina Gil \textit{et al.} [NA62],
JHEP \textbf{06}, 093 (2021)
doi:10.1007/JHEP06(2021)093
[arXiv:2103.15389 [hep-ex]].

\bibitem{KOTO:2018dsc}
J.~K.~Ahn \textit{et al.} [KOTO],
Phys. Rev. Lett. \textbf{122}, no.2, 021802 (2019)
doi:10.1103/PhysRevLett.122.021802
[arXiv:1810.09655 [hep-ex]].

\bibitem{Buchalla:1993bv}
G.~Buchalla and A.~J.~Buras,
Nucl. Phys. B \textbf{400}, 225-239 (1993)
doi:10.1016/0550-3213(93)90405-E

\bibitem{Buchalla:1993wq}
G.~Buchalla and A.~J.~Buras,
Nucl. Phys. B \textbf{412}, 106-142 (1994)
doi:10.1016/0550-3213(94)90496-0
[arXiv:hep-ph/9308272 [hep-ph]].

\bibitem{Misiak:1999yg}
M.~Misiak and J.~Urban,
Phys. Lett. B \textbf{451}, 161-169 (1999)
doi:10.1016/S0370-2693(99)00150-1
[arXiv:hep-ph/9901278 [hep-ph]].

\bibitem{Buras:2005gr}
A.~J.~Buras, M.~Gorbahn, U.~Haisch and U.~Nierste,
Phys. Rev. Lett. \textbf{95}, 261805 (2005)
doi:10.1103/PhysRevLett.95.261805
[arXiv:hep-ph/0508165 [hep-ph]].

\bibitem{Buras:2006gb}
A.~J.~Buras, M.~Gorbahn, U.~Haisch and U.~Nierste,
JHEP \textbf{11}, 002 (2006)
[erratum: JHEP \textbf{11}, 167 (2012)]
doi:10.1007/JHEP11(2012)167
[arXiv:hep-ph/0603079 [hep-ph]].

\bibitem{Gorbahn:2004my}
M.~Gorbahn and U.~Haisch,
Nucl. Phys. B \textbf{713}, 291-332 (2005)
doi:10.1016/j.nuclphysb.2005.01.047
[arXiv:hep-ph/0411071 [hep-ph]].

\bibitem{Mescia:2007kn}
F.~Mescia and C.~Smith,
Phys. Rev. D \textbf{76}, 034017 (2007)
doi:10.1103/PhysRevD.76.034017
[arXiv:0705.2025 [hep-ph]].

\bibitem{Isidori:2005xm}
G.~Isidori, F.~Mescia and C.~Smith,
Nucl. Phys. B \textbf{718}, 319-338 (2005)
doi:10.1016/j.nuclphysb.2005.04.008
[arXiv:hep-ph/0503107 [hep-ph]].

\bibitem{Bordone:2021oof}
M.~Bordone, B.~Capdevila and P.~Gambino,
Phys. Lett. B \textbf{822}, 136679 (2021)
doi:10.1016/j.physletb.2021.136679
[arXiv:2107.00604 [hep-ph]].

\bibitem{Aoki:2021kgd}
Y.~Aoki \textit{et al.} [Flavour Lattice Averaging Group (FLAG)],
Eur. Phys. J. C \textbf{82}, no.10, 869 (2022)
doi:10.1140/epjc/s10052-022-10536-1
[arXiv:2111.09849 [hep-lat]].

\bibitem{Buras:2021nns}
A.~J.~Buras and E.~Venturini,
Acta Phys. Polon. B \textbf{53}, no.6, A1
doi:10.5506/APhysPolB.53.6-A1
[arXiv:2109.11032 [hep-ph]].

\bibitem{Buras:2022nfn}
A.~J.~Buras,
Eur. Phys. J. C \textbf{82}, no.7, 612 (2022)
doi:10.1140/epjc/s10052-022-10566-9
[arXiv:2204.10337 [hep-ph]].

\bibitem{Buras:2022qip}
A.~J.~Buras,
[arXiv:2209.03968 [hep-ph]].

\bibitem{Buras:2022wpw}
A.~J.~Buras and E.~Venturini,
Eur. Phys. J. C \textbf{82}, no.7, 615 (2022)
doi:10.1140/epjc/s10052-022-10583-8
[arXiv:2203.11960 [hep-ph]].

\bibitem{LHCb:2020ycd}
R.~Aaij \textit{et al.} [LHCb],
Phys. Rev. Lett. \textbf{125}, no.23, 231801 (2020)
doi:10.1103/PhysRevLett.125.231801
[arXiv:2001.10354 [hep-ex]].

\bibitem{KTeV:2003sls}
A.~Alavi-Harati \textit{et al.} [KTeV],
Phys. Rev. Lett. \textbf{93}, 021805 (2004)
doi:10.1103/PhysRevLett.93.021805
[arXiv:hep-ex/0309072 [hep-ex]].

\bibitem{KTEV:2000ngj}
A.~Alavi-Harati \textit{et al.} [KTEV],
Phys. Rev. Lett. \textbf{84}, 5279-5282 (2000)
doi:10.1103/PhysRevLett.84.5279
[arXiv:hep-ex/0001006 [hep-ex]].

\bibitem{Isidori:2003ts}
G.~Isidori and R.~Unterdorfer,
JHEP \textbf{01}, 009 (2004)
doi:10.1088/1126-6708/2004/01/009
[arXiv:hep-ph/0311084 [hep-ph]].

\bibitem{DAmbrosio:2017klp}
G.~D'Ambrosio and T.~Kitahara,
Phys. Rev. Lett. \textbf{119}, no.20, 201802 (2017)
doi:10.1103/PhysRevLett.119.201802
[arXiv:1707.06999 [hep-ph]].

\bibitem{Mescia:2006jd}
F.~Mescia, C.~Smith and S.~Trine,
JHEP \textbf{08}, 088 (2006)
doi:10.1088/1126-6708/2006/08/088
[arXiv:hep-ph/0606081 [hep-ph]].

\bibitem{Dery:2021mct}
A.~Dery, M.~Ghosh, Y.~Grossman and S.~Schacht,
JHEP \textbf{07}, 103 (2021)
doi:10.1007/JHEP07(2021)103
[arXiv:2104.06427 [hep-ph]].

\bibitem{Aebischer:2020dsw}
J.~Aebischer, C.~Bobeth, A.~J.~Buras and J.~Kumar,
JHEP \textbf{12}, 187 (2020)
doi:10.1007/JHEP12(2020)187
[arXiv:2009.07276 [hep-ph]].

\bibitem{Aebischer:2018csl}
J.~Aebischer, C.~Bobeth, A.~J.~Buras and D.~M.~Straub,
Eur. Phys. J. C \textbf{79}, no.3, 219 (2019)
doi:10.1140/epjc/s10052-019-6715-6
[arXiv:1808.00466 [hep-ph]].

\bibitem{Aebischer:2018quc}
J.~Aebischer, C.~Bobeth, A.~J.~Buras, J.~M.~G\'erard and D.~M.~Straub,
Phys. Lett. B \textbf{792}, 465-469 (2019)
doi:10.1016/j.physletb.2019.04.016
[arXiv:1807.02520 [hep-ph]].

\bibitem{Aebischer:2018rrz}
J.~Aebischer, A.~J.~Buras and J.~M.~G\'erard,
JHEP \textbf{02}, 021 (2019)
doi:10.1007/JHEP02(2019)021
[arXiv:1807.01709 [hep-ph]].

\bibitem{Aebischer:2021hws}
J.~Aebischer, C.~Bobeth, A.~J.~Buras and J.~Kumar,
JHEP \textbf{12}, 043 (2021)
doi:10.1007/JHEP12(2021)043
[arXiv:2107.12391 [hep-ph]].

\bibitem{Aebischer:2019blw}
J.~Aebischer, A.~J.~Buras, M.~Cerd\`a-Sevilla and F.~De Fazio,
JHEP \textbf{02}, 183 (2020)
doi:10.1007/JHEP02(2020)183
[arXiv:1912.09308 [hep-ph]].

\bibitem{Aebischer:2020mkv}
J.~Aebischer, A.~J.~Buras and J.~Kumar,
JHEP \textbf{12}, 097 (2020)
doi:10.1007/JHEP12(2020)097
[arXiv:2006.01138 [hep-ph]].

\bibitem{Aebischer:2020lsx}
J.~Aebischer and J.~Kumar,
JHEP \textbf{09}, 187 (2020)
doi:10.1007/JHEP09(2020)187
[arXiv:2005.12283 [hep-ph]].

\bibitem{Crosas:2022quq}
\`O.~L.~Crosas, G.~Isidori, J.~M.~Lizana, N.~Selimovic and B.~A.~Stefanek,
[arXiv:2207.00018 [hep-ph]].

\bibitem{Li:2019fhz}
T.~Li, X.~D.~Ma and M.~A.~Schmidt,
Phys. Rev. D \textbf{101}, no.5, 055019 (2020)
doi:10.1103/PhysRevD.101.055019
[arXiv:1912.10433 [hep-ph]].

\bibitem{Deppisch:2020oyx}
F.~F.~Deppisch, K.~Fridell and J.~Harz,
JHEP \textbf{12}, 186 (2020)
doi:10.1007/JHEP12(2020)186
[arXiv:2009.04494 [hep-ph]].


\end{thebibliography}
\end{document}